\documentclass[12pt]{article}
\usepackage{a4wide}
\usepackage{latexsym}
\usepackage{cite}
\usepackage{axodraw}
\usepackage{amssymb}
\usepackage{epsfig} 

\usepackage[nice]{nicefrac}

\usepackage{pslatex}
\usepackage[latin1]{inputenc}
\usepackage[T1]{fontenc}

\newcommand{\bq}{\begin{eqnarray}}
\newcommand{\eq}{\end{eqnarray}}
\renewcommand{\l}{\langle}
\renewcommand{\r}{\rangle} 
\newcommand{\eps}{\varepsilon}

\begin{document}

\thispagestyle{empty}

\begin{flushright}
  MZ-TH/06-03 \\
\end{flushright}

\vspace{1.5cm}

\begin{center}
  {\Large\bf A comparison of efficient methods for the computation of Born gluon amplitudes\\
  }
  \vspace{1cm}
  {\large Michael Dinsdale, Marko Ternick and Stefan Weinzierl\\
  \vspace{1cm}
      {\small \em Institut f{\"u}r Physik, Universit{\"a}t Mainz,}\\
      {\small \em D - 55099 Mainz, Germany}\\
  } 
\end{center}

\vspace{2cm}

\begin{abstract}\noindent
  {
We compare four different methods for the numerical computation of the 
pure gluonic amplitudes in the Born approximation.
We are in particular interested in the efficiency of the various methods as
the number $n$ of the external particles increases.
In addition we investigate the numerical accuracy in critical phase space regions.
The methods considered are based on
(i) Berends-Giele recurrence relations,
(ii) scalar diagrams,
(iii) MHV vertices and
(iv) BCF recursion relations.
   }
\end{abstract}

\vspace*{\fill}

\newpage

\section{Introduction}
\label{sect:intro}

The fast and accurate computation of multi-parton amplitudes in QCD 
is essential for our understanding of multi-jet processes at the LHC.
It is a well-known fact that the conventional approach -- summing up all Feynman diagrams --
already reaches its limit for Born amplitudes when the number of external partons is about $5$.
More efficient methods construct the amplitudes recursively from smaller building blocks.
The recursive approach allows to recycle information from already calculated pieces.
An example of these methods are the Berends-Giele recurrence 
relations~\cite{Berends:1987me,Kosower:1989xy,Caravaglios:1995cd,Kanaki:2000ey,Moretti:2001zz}.
The basic building blocks here are currents with one parton off-shell.

Recently, a number of new methods for the calculation of
helicity amplitudes in QCD have been introduced, motivated by the
relationship of QCD amplitudes to twistor string
theory~\cite{Witten:2003nn}.  In the Cachazo~-~Svr\v{c}ek~-~Witten
(CSW) construction~\cite{Cachazo:2004kj,Risager:2005vk}, tree level QCD amplitudes
are constructed from vertices that are off-shell continuations of
maximal-helicity-violating (MHV) amplitudes~\cite{Parke:1986gb},
connected by scalar propagators.  
Subsequently a set of recursion
relations has been found~\cite{Britto:2004ap,Britto:2005fq,Draggiotis:2005wq,Vaman:2005dt} that
involve only on-shell amplitudes with shifted, complex external
momenta.  
In~\cite{Schwinn:2005pi,Schwinn:2005zm} a method has been presented, which is close in spirit to the 
Berends-Giele recursion relations, but which involves only a set of primitive vertices and scalar propagators.

These new methods enriched our understanding of the structure of QCD amplitudes.
In particular they gave a precise answer to the question of what complexity we should expect in the result for a 
particular helicity amplitude, if we go beyond the simple maximal-helicity-violating ones.
As an application, the singular behaviour of tree amplitudes in the multi-collinear limit 
could be derived from these methods \cite{Birthwright:2005ak,Birthwright:2005vi}.
Furthermore these new methods turn out to be very useful in the construction of 
one-loop amplitudes \cite{Cachazo:2004zb,Britto:2004nc,Bena:2004xu,Bern:2004ky,Bidder:2004tx,Bern:2004bt,Bern:2005hs,Bidder:2004vx,Bidder:2005ri,Brandhuber:2004yw,Bedford:2004py,Quigley:2004pw,Roiban:2004ix,Luo:2004ss,Luo:2004nw}
in conjunction with the unitary-based method \cite{Bern:1994zx,Bern:1995cg}.

Besides this undeniable progress in the understanding of the analytical structure of QCD amplitudes, one question
immediately arises: Do the new methods lead to improved algorithms for the computation of Born amplitudes
in a purely numerical approach ?
To answer this question we examine in this paper four different methods for the numerical computation of the 
pure gluonic amplitudes in the Born approximation.
The methods considered are based on
(i) Berends-Giele recurrence relations,
(ii) scalar diagrams,
(iii) MHV vertices and
(iv) BCF recursion relations.
We compare the efficiency of the various methods as
the number $n$ of external particles increases.
In addition we investigate the numerical accuracy in critical phase space regions.

This article is organised as follows:
In section~\ref{sect:methods} we present the four different methods for the computation of the pure gluonic amplitudes
in the Born approximation.
In section~\ref{sect:numeric} we compare the performance of the various methods and study the numerical stability
in critical phase space regions.
Section~\ref{sect:conclusions} contains our conclusions.
Technical details are collected in the appendix.
Appendices~\ref{appendix:spinors} and ~\ref{appendix:Feynman}
define our conventions for spinors and Feynman rules.
Appendix \ref{appendix:splitting functions} collects the $g \rightarrow g g$ splitting functions,
which describe the collinear limit.
In appendix~\ref{appendix:optimization} we comment on various optimisation techniques, which can be used to speed up
the computation.


\section{Description of the different methods}
\label{sect:methods}

The tree level amplitude with $n$ external gluons may be written in 
the form
\bq
{\cal A}_n(k_1^{\lambda_1},...,k_n^{\lambda_n}) & = & g^{n-2} \sum\limits_{\sigma \in S_{n}/Z_{n}} 2 \; \mbox{Tr} \left(
 T^{a_{\sigma(1)}} ... T^{a_{\sigma(n)}} \right)
 A_{n}\left( k_{\sigma(1)}^{\lambda_{\sigma(1)}}, ..., k_{\sigma(n)}^{\lambda_{\sigma(n)}} \right), 
\eq
where the sum runs over all non-cyclic permutations of the external gluon legs.
The symbol $k_j$ denotes the four-momentum of the $j$-th gluon and $\lambda_j$ its helicity.
$g$ denotes the strong coupling constant and $T^a$ the colour matrices, which are normalised such that
$\mbox{Tr}(T^a T^b) = \nicefrac{1}{2}\,\, \delta^{ab}$. 
The quantities $A_n$, called the partial amplitudes, contain the 
kinematic information.
They are colour-ordered, i.e. only diagrams with a particular cyclic ordering of the gluons 
contribute.

In the computation of observables and cross-sections, the amplitude squared enters, 
summed over all helicities and colour degrees of freedom.
We have
\bq
\label{amplitude_squared}
 \left| {\cal A}_n \right|^2 & = &
 2^{2-n} g^{2n-4} N_c^n 
 \sum\limits_{\lambda_1,...,\lambda_n}
 \sum\limits_{\sigma \in S_{n}/Z_{n}} 
 \left| A_{n}\left( k_{\sigma(1)}^{\lambda_{\sigma(1)}}, ..., k_{\sigma(n)}^{\lambda_{\sigma(n)}} \right)
 \right|^2
 + {\cal O}\left(\frac{1}{N_c^2} \right),
\eq
where $N_c=3$ denotes the number of colours.
The colour-suppressed terms consist of interference terms between partial amplitudes with different colour-orderings.

In this paper we investigate various methods for the efficient computation of the partial amplitudes
$A_n$.
We would like to separate this issue from the sum over all colour structures implicit in eq.~(\ref{amplitude_squared}).
Therefore we focus on the quantity
\bq
\label{benchmark_M}
 {\cal M}_n & = &
 \sum\limits_{\lambda_1,...,\lambda_n}
 \left| A_{n}\left( k_{1}^{\lambda_{1}}, ..., k_{n}^{\lambda_{n}} \right)
 \right|^2.
 \eq
${\cal M}_n$ gives the leading-colour contribution to eq.~(\ref{amplitude_squared}), but in the context here it should
be regarded as a quantity which depends only on the kinematical information and which helps to study the efficiency of
various methods to calculate the partial amplitudes $A_n$.
It should be noted that only half of the helicity configurations need to be calculated, since parity relates a partial
amplitude to the one with all helicities reversed:
\bq
A_{n}\left( k_{1}^{\lambda_{1}}, ..., k_{n}^{\lambda_{n}} \right)
 & = & - A_{n}\left( k_{1}^{-\lambda_{1}}, ..., k_{n}^{-\lambda_{n}} \right)^\ast.
\eq
In addition we investigate the numerical accuracy of the various methods in critical phase space regions.
These are regions where one or more partons become unresolved.
The simplest case involves single unresolved configurations, where one parton becomes either soft or collinear to a second
parton.
In the limit where one gluon $j$ becomes soft, the partial amplitudes behave as
\bq
A_{n+1}(k_1,...k_j^+,...,k_{n+1}) & \stackrel{k_j \; soft}{\longrightarrow} & 
  \sqrt{2} \frac{\l k_{j-1} k_{j+1} \r}{\l k_{j-1} k_j \r \l k_j k_{j+1} \r} 
                              A_n(k_1,...,k_{n+1}), 
 \nonumber \\
A_{n+1}(k_1,...k_j^-,...,k_{n+1}) & \stackrel{k_j \; soft}{\longrightarrow} & 
  \sqrt{2} \frac{[ k_{j+1} k_{j-1} ]}{[ k_{j+1} k_j ] [ k_j k_{j-1} ]} 
                              A_n(k_1,...,k_{n+1}).
\eq
$\l k_i k_j \r$ and $[k_i k_j ]$ denote spinor products, which are defined in appendix~\ref{appendix:spinors}. 
The quantity ${\cal M}_n$ factorises in the soft limit as
\bq
\label{eq:soft_singularity}
{\cal M}_{n+1}(k_1,...k_j,...,k_{n+1}) & \stackrel{k_j \; soft}{\longrightarrow} & 
  2 \frac{(2 k_{j-1} k_{j+1})}{(2 k_{j-1} k_j) (2 k_j k_{j+1})} 
                              {\cal M}_n(k_1,...,k_{n+1}).
\eq
In the collinear limit tree-level partial amplitudes factorise according to
\bq
A_{n+1}(...,k_a,k_b,...) \stackrel{k_a || k_b}{\longrightarrow} \sum\limits_{\lambda = \pm} \mbox{Split}_{-\lambda} 
(k_a^{\lambda_a},k_b^{\lambda_b}) A_n(...,K^\lambda,...),
\eq
where $k_a$ and $k_b$ are the momenta of two adjacent legs.
In the collinear limit we have
$K = k_a + k_b$, $k_a = z K$ and $k_b = (1-z) K$. $\lambda$, $\lambda_a$ and $\lambda_b$ denote the
corresponding helicities.
The splitting functions are listed in appendix \ref{appendix:splitting functions}.
In the collinear limit the quantity ${\cal M}_n$ behaves as \cite{Catani:1997vz,Weinzierl:1999yf,Weinzierl:2005dd}
\bq
\label{eq:coll_singularity}
{\cal M}_{n+1}(...,k_a,k_b,...) 
 \stackrel{k_a || k_b}{\longrightarrow} 
 \frac{2}{2k_a k_b} \left( \frac{2}{1-z}+\frac{2}{z}-4 \right) {\cal M}_n(...,K^\lambda,...)
 +\frac{8}{(2k_ak_b)^2} {\cal S}_n,
\eq
where the spin-correlation is given by
\bq
 {\cal S}_n & = &
 \sum\limits_{\lambda_1,...,\lambda_{a-1}, \lambda_{b+1},...,\lambda_n}
 \left| 
        E A_{n}\left( k_{1}^{\lambda_{1}}, ..., K^+, ..., k_{n}^{\lambda_{n}} \right)
      + E^\ast A_{n}\left( k_{1}^{\lambda_{1}}, ..., K^-, ..., k_{n}^{\lambda_{n}} \right)
 \right|^2
 \eq
and
\bq 
 E & = & z \frac{\l k_b+ | k_a | K+\r}{\sqrt{2} [ K k_b]}.
\eq


\subsection{Berends-Giele type recurrence relations}

Berends-Giele type recurrence relations \cite{Berends:1987me,Kosower:1989xy}
build partial amplitudes from smaller building blocks, usually
called colour-ordered off-shell currents.
Off-shell currents are objects with $n$ on-shell legs and one additional leg off-shell.
Momentum conservation is satisfied. It should be noted that
off-shell currents are not gauge-invariant objects.
Recurrence relations relate off-shell currents with $n$ legs 
to off-shell currents with fewer legs.
\\
\\
The recursion starts with $n=1$:
\bq
J^\mu(k_1) & = & \eps^\mu(k_1,q).
\eq
$\eps^\mu$ is the polarisation vector of the gluon and $q$ an arbitrary light-like reference momentum.
We have the explicit formulae
\bq
\label{eq:pol}
\eps_{\mu}^{+}(k,q) = \frac{\langle q-|\gamma_{\mu}|k-\rangle}{\sqrt{2} \langle q- | k + \rangle},
 & &
\eps_{\mu}^{-}(k,q) = \frac{\langle q+|\gamma_{\mu}|k+\rangle}{\sqrt{2} \langle k + | q - \rangle}.
\eq
The recursive relation states that a gluon couples to other gluons only via the three- or four-gluon
vertices :
\bq
\label{Berends_Giele_recursion}
 J^\mu(k_1^{\lambda_1},...,k_n^{\lambda_n}) & = & 
 \frac{-i}{K^2_{1,n}} 
 \left[ 
        \sum\limits_{j=1}^{n-1} V_3^{\mu\nu\rho}(-K_{1,n},K_{1,j},K_{j+1,n})
                                J_\nu(k_1^{\lambda_1},...,k_j^{\lambda_j}) J_\rho(k_{j+1}^{\lambda_{j+1}},...,k_n^{\lambda_n}) 
 \right. \nonumber \\
 & & \left. 
        + \sum\limits_{j=1}^{n-2} \sum\limits_{l=j+1}^{n-1} V_4^{\mu\nu\rho\sigma} 
            J_\nu(k_1^{\lambda_1},...,k_j^{\lambda_j}) J_\rho(k_{j+1}^{\lambda_{j+1}},...,k_l^{\lambda_l}) J_\sigma(k_{l+1}^{\lambda_{l+1}},...,k_n^{\lambda_n}) 
     \right],
\eq
where
\bq
K_{i,j} & = & k_i + k_{i+1} + ... + k_j 
\eq
and $V_3$ and $V_4$ are the colour-ordered three-gluon and four-gluon vertices
\bq
\label{Feynman_rules}
 V_3^{\mu\nu\rho}(k_1,k_2,k_3) 
 & = & 
 i \left[
          g^{\mu\nu} \left( k_1^\rho - k_2^\rho \right)
        + g^{\nu\rho} \left( k_2^\mu - k_3^\mu \right)
        + g^{\rho\mu} \left( k_3^\nu - k_1^\nu \right)
   \right],
 \nonumber \\
 V_4^{\mu\nu\rho\sigma} & = & i \left( 2 g^{\mu\rho} g^{\nu\sigma} - g^{\mu\nu} g^{\rho\sigma} -g^{\mu\sigma} g^{\nu\rho} \right).
\eq
The gluon current $J_\mu$ is conserved:
\bq
\left( \sum\limits_{i=1}^n k_i^\mu \right) J_\mu & = & 0.
\eq
Therefore terms proportional to $K_{1,j}^\nu$ and proportional to $K_{j+1,n}^\rho$ can be dropped 
in eq. (\ref{Berends_Giele_recursion}) and, using momentum conservation, it is legitimate to use the slightly simpler expression
\bq
 V_3^{\mu\nu\rho}(k_1,k_2,k_3) 
 & = & 
   i \left( g^{\nu\rho} (k_2-k_3)^\mu + 2 g^{\rho\mu} k_3^\nu - 2 g^{\mu\nu} k_2^\rho \right).
\eq
for the three gluon vertex in eq. (\ref{Berends_Giele_recursion}).
\\
The partial amplitude $A_n(k_1^{\lambda_1},...,k_n^{\lambda_n})$ is obtained from the 
gluonic current $J^\mu(k_1^{\lambda_1},...,k_{n-1}^{\lambda_{n-1}})$
by multiplying by the inverse gluon propagator and contracting with the polarisation vector for gluon $n$:
\bq
 A_n(k_1^{\lambda_1},...,k_n^{\lambda_n}) & = & \eps_\mu^{\lambda_n}(k_n,q) \cdot \left(i K_{1,n-1}^2 \right) 
                                                 J^\mu(k_1^{\lambda_1},...,k_{n-1}^{\lambda_{n-1}}).
\eq
A close inspection of the recursion relation eq.~(\ref{Berends_Giele_recursion}) shows 
that only the quantities $J^\mu(k_i^{\lambda_i},...,k_{j-1}^{\lambda_{j-1}})$ which respect the original order
need to be calculated.
Therefore an efficient implementation stores a list of four-momenta
\bq
 \left[ k_1, k_2, ..., k_n \right]
\eq
and a list of helicities 
\bq
 \left[ \lambda_1, \lambda_2, ..., \lambda_n \right]
\eq
in memory and passes to the subroutine just two integers $i$ and $j$, indicating that the quantity
\bq
 J^\mu(k_i^{\lambda_i},...,k_{j-1}^{\lambda_{j-1}})
\eq
should be computed.


\subsection{Recursive calculation with scalar diagrams}

A modification of the Berends-Giele recursion relation was advocated in refs. \cite{Kosower:1989xy} and \cite{Schwinn:2005pi}.
In this approach all summations over Lorentz indices are replaced by a sum over the two physical polarisations.
This reduces the number of multiplications needed for a contraction from four to two.
The resulting recurrence relation consists of scalar propagators and a set of primitive vertices.

Let $q$ be a null-vector, which will be kept fixed throughout the discussion.
Using $q$, any massive vector $k$ can be written as 
a sum of two null-vectors $k^\flat$ and $q$ \cite{Kosower:2004yz}:
\bq
\label{offshellcont}
k & = & k^\flat + \frac{k^2}{2kq} q.
\eq
Obviously, if $k^2=0$, we have $k = k^\flat$.
Note further that $2kq = 2k^\flat q$.
Using eq. (\ref{offshellcont}) we may associate a massless four-vector $k^\flat$
to any four-vector $k$.
Using the projection onto $k^\flat$ we define the off-shell
continuation of Weyl spinors as
\bq
\label{offshellcontspinor}
 | k \pm \r & \rightarrow & | k^\flat \pm \r,
 \nonumber \\
 \l k \pm | & \rightarrow & \l k^\flat \pm |.
\eq
We are going to use the following abbreviations:
\bq
 & &
 \left\l i j \right\r = \left\l k_i^\flat - | k_j^\flat + \right\r,
  \;\;\;\;\;\;
 \left[ i j \right] = \left\l k_i^\flat + | k_j^\flat - \right\r,
 \nonumber \\
 & &
 \left\l i- \left| j \pm k \right| l- \right\r =
 \left\l k^\flat_i- \left| k\!\!\!/^\flat_j \pm k\!\!\!/^\flat_k \right| k^\flat_l- \right\r. 
\eq
In spinor products, the projections $k^\flat$ are always used.
Let us define an ``off-shell amplitude''
\bq
 O_n\left(k_1^{\lambda_1}, k_2^{\lambda_2}, ..., k_n^{\lambda_n}\right),
\eq
depending on $n$ external momenta $k_i$ and helicities $\lambda_i$. The momenta need not be on-shell,
but momentum conservation is imposed:
\bq
 \sum\limits_{j=1}^n k_j 
 & = & 0.
\eq
By definition,
the off-shell amplitudes $O_n$ are calculated from all Feynman diagrams contributing to the
cyclic-ordered partial amplitude $A_n$, by using the off-shell
continuation eq.~(\ref{offshellcontspinor}) 
for all external spinors and polarisation vectors, and by using the axial gauge
for all internal gluon propagators.
Compared to off-shell currents, which are used in recurrence 
relations of Berends-Giele type, an off-shell amplitude may have more 
than one leg off-shell.
By construction, if all external particles are on-shell, the off-shell
amplitude $O_n$ coincides with the physical amplitude $A_n$.
We have the following recurrence relation:
\bq
\label{recursion_scalar}
\lefteqn{
 O_n\left( k_1^{\lambda_1}, ..., k_n^{\lambda_n} \right)
 =  
 \sum\limits_{\lambda, \lambda' = \pm}
 \sum\limits_{j=2}^{n-1} V_3( K_{1,j-1}^\lambda, K_{j,n-1}^{\lambda'}, k_n^{\lambda_n})
} & &
 \\
 & &
 \times 
 \frac{i}{K_{1,j-1}^2} O_j(k_1^{\lambda_1},...,k_{j-1}^{\lambda_{j-1}}, (-K_{1,j-1})^{-\lambda} )
 \frac{i}{K_{j,n-1}^2} O_{n-j+1}(k_j^{\lambda_j},...,k_{n-1}^{\lambda_{n-1}}, (-K_{j,n-1})^{-\lambda'} )
 \nonumber \\
 & &
 +  \sum\limits_{\lambda, \lambda', \lambda'' = \pm}
 \sum\limits_{j=2}^{n-2} \sum\limits_{l=j+1}^{n-1}
 V_4\left( K_{1,j-1}^\lambda, K_{j,l-1}^{\lambda'}, K_{l,n-1}^{\lambda''}, k_n^{\lambda_n} \right)
 \frac{i}{K_{1,j-1}^2} O_j(k_1^{\lambda_1},...,k_{j-1}^{\lambda_{j-1}}, (-K_{1,j-1})^{-\lambda} )
 \nonumber \\
 & &
 \times
 \frac{i}{K_{j,l-1}^2} O_{l-j+1}(k_j^{\lambda_j},...,k_{l-1}^{\lambda_{l-1}}, (-K_{j,l-1})^{-\lambda'} )
 \frac{i}{K_{l,n-1}^2} O_{n-l+1}(k_l^{\lambda_l},...,k_{n-1}^{\lambda_{n-1}}, (-K_{l,n-1})^{-\lambda''} ),
 \nonumber 
\eq
where we define the two-point amplitude to be the inverse propagator:
\bq
 O_2(k_j^\pm, -K_{j,j}^\mp ) & = & - i k_j^2.
\eq 
The partial amplitude $A_n$ coincides with $O_n$,
if all gluons are on-shell:
\bq
 A_n(k_1^{\lambda_1},...,k_n^{\lambda_n}) & = &
 O_n(k_1^{\lambda_1},...,k_n^{\lambda_n}).
\eq
There is only a limited number of non-zero vertices, which are listed in appendix \ref{appendix:scalar_rules}.
This allows for a high degree of optimisation in the calculation of these vertices.
The double and triple sums over the intermediate helicities in eq.~(\ref{recursion_scalar})
reduce in all cases to three non-vanishing terms.

On the other hand it should be pointed out that in this approach the four-valent vertices depend 
(as do the three-valent vertices) on the momenta attached to these vertices.
This should be compared to the standard Feynman rules, which enter the Berends-Giele recurrence relation,
where the four-gluon vertex in eq.~(\ref{Feynman_rules}) is independent of the momenta.

As in the Berends-Giele recurrence relation, an efficient implementation stores the sequence of four-momenta and helicities
in a central place and just passes two integers $i$ and $j$ to the implementation of the recurrence relation,
indicating the starting and ending points.


\subsection{Recursive calculation with MHV vertices}

In the Cachazo~-~Svr\v{c}ek~-~Witten
(CSW) construction~\cite{Cachazo:2004kj}, tree level QCD amplitudes
are constructed from vertices that are off-shell continuations of
maximal-helicity-violating (MHV) amplitudes, connected by scalar propagators.  
In maximal-helicity-violating amplitudes all gluons 
except two have the same helicity.
Compact formulae for these amplitudes have been known for a long time~\cite{Parke:1986gb}.
Using the off-shell continuation eq. (\ref{offshellcontspinor})
the MHV-amplitudes serve as new vertices:
\bq
\label{Parke_Taylor}
V_n(k_1^+,...,k_j^-,...,k_k^-,...,k_n^+) 
 & = & i \left( \sqrt{2} \right)^{n-2} 
 \frac{\langle j k \rangle^4}{\langle 1 2 \rangle ... \langle n 1 \rangle},
 \nonumber \\
V_n(k_1^-,...,k_j^+,...,k_k^+,...,k_n^-) 
 & = & i \left( \sqrt{2} \right)^{n-2} 
 \frac{[ k j ]^4}{[1 n ] [n (n-1)] ... [ 2 1 ]}.
\eq
Each MHV vertex has exactly two lines carrying negative helicity and at least one line carrying positive helicity.

Bena, Bern and Kosower~\cite{Bena:2004ry} derived a recursive formulation, which allows to obtain vertices
with more gluons of negative helicity from simpler building blocks:
\bq
\label{MHV_recursion}
\lefteqn{
 V_n(k_1^{\lambda_1},\ldots,k_n^{\lambda_n}) = 
{1\over (n_{neg}-2)}\sum_{j=1}^n\sum_{l=j+1}^{j-3}
{i\over K_{j,l}^2} 
 V_{(l-j+2) \;\mbox{\scriptsize mod}\; n}(k_{j}^{\lambda_{j}},\ldots,k_{l}^{\lambda_{l}},(-K_{j,l})^{-}) 
}
 & & \nonumber \\
 & & 
 \hspace*{45mm}
 \times
 V_{(j-l) \;\mbox{\scriptsize mod}\; n}(k_{l+1}^{\lambda_{l+1}},\ldots,k_{j-1}^{\lambda_{j-1}},(-K_{(l+1),(j-1)})^+),
\hspace*{5mm}
\eq
where $n_{neg}$ is the number of negative helicity gluons. 
The recursion starts if $n_{neg}$ is less than two. For $n_{neg}=0$ or $n_{neg}=1$ 
the quantity $V_n(k_1^{\lambda_1},\ldots,k_n^{\lambda_n})$ vanishes. For $n_{neg}=2$ it is given by eq. (\ref{Parke_Taylor}).
Again, the partial amplitude $A_n$ coincides with $V_n$,
if all gluons are on-shell:
\bq
 A_n(k_1^{\lambda_1},...,k_n^{\lambda_n}) & = &
 V_n(k_1^{\lambda_1},...,k_n^{\lambda_n}).
\eq
There are two points which should be noted:
First of all, there is a double sum in eq.~(\ref{MHV_recursion}), which over-counts each contribution $(n_{neg}-2)$ times.
This over-counting is compensated by the explicit factor $1/(n_{neg}-2)$ in front.
\\
Secondly, it is no longer possible to work with a static list of four-vectors and helicities, as was the case for the first
two methods.
The recursion relation eq.~(\ref{MHV_recursion}) inserts the four-momenta
$-K_{j,l}$ and $-K_{(l+1),(j-1)}$ into the cyclic order. Therefore the lists of momenta and helicities
have to be updated at each step of the recursion.
This is best implemented by a double-linked list, which allows for the insertion of the new elements without
copying the remaining ones.


\subsection{Recursive calculation with shifted momenta}

Britto, Cachazo and Feng \cite{Britto:2004ap}
gave a recursion relation for the calculation of the $n$-gluon amplitude
which involves only on-shell amplitudes.
To describe this method it is best not to view the partial amplitude $A_n$ as a function
of the four-momenta $k_j^\mu$, but to replace each four-vector by a pair of two-component
Weyl spinors.
\\
In detail this is done as follows:
Each four-vector $K_\mu$ has a bispinor representation, given by
\bq
\label{bispinor_representation}
 K_{A\dot{B}} = K_\mu \sigma^\mu_{A\dot{B}},
 & &
 K_\mu = \frac{1}{2} K_{A\dot{B}} \bar{\sigma}_\mu^{\dot{B}A}.
\eq
For null-vectors this bispinor representation factorises into a dyad of Weyl spinors:
\bq
\label{dyad}
 k_\mu k^\mu = 0
 & \Leftrightarrow &
 k_{A\dot{B}} = k_{A} k_{\dot{B}}.
\eq
The equations~(\ref{bispinor_representation}) and~(\ref{dyad})
allow us to convert any light-like four-vector into a dyad of Weyl spinors and vice versa.
Therefore the partial amplitude $A_n$, being originally a function of the momenta $k_j$ and helicities
$\lambda_j$,
can equally be viewed as a function of the Weyl spinors $k_A^j$, $k_{\dot{B}}^j$ and the helicities
$\lambda_j$:
\bq
 A_n(k_1^{\lambda_1},...,k_n^{\lambda_n}) & = & 
 A_n( k_A^1, k_{\dot{B}}^1, \lambda_1, ..., k_A^n, k_{\dot{B}}^n, \lambda_n).
\eq
Note that for an arbitrary pair of Weyl spinors, the corresponding four-vector will in general be complex-valued.
If $(\lambda_n,\lambda_1) \neq (+,-)$ we have the following recurrence relation:
\bq
\label{on_shell_recursion}
\lefteqn{
A_n\left( k_A^1, k_{\dot{B}}^1, \lambda_1, ..., k_A^n, k_{\dot{B}}^n, \lambda_n\right)
 = 
 } & & \\
 & & 
 \hspace*{20mm}
 \sum\limits_{j=3}^{n-1} \sum\limits_{\lambda=\pm}
  A_{j}\left( \hat{k}_A^1, k_{\dot{B}}^1, \lambda_1, 
              k_A^2, k_{\dot{B}}^2, \lambda_2, 
              ..., 
	      k_A^{j-1}, k_{\dot{B}}^{j-1}, \lambda_{j-1},
              i \hat{K}_A, i \hat{K}_{\dot{B}}, -\lambda
              \right)
 \nonumber \\
 & &
 \hspace*{20mm}
  \times
  \frac{i}{K_{1,j-1}^2} 
  A_{n-j+2}\left( 
                  \hat{K}_A, \hat{K}_{\dot{B}}, \lambda,
                  k_A^j, k_{\dot{B}}^j, \lambda_j, 
                  ..., 
                  k_A^{n-1}, k_{\dot{B}}^{n-1}, \lambda_{n-1},
                  k_A^n, \hat{k}_{\dot{B}}^n, \lambda_n \right).
 \nonumber 
\eq
If $(\lambda_n,\lambda_1) = (+,-)$ we can always cyclic permute the arguments, such that
$(\lambda_n,\lambda_1) \neq (+,-)$.
This is possible, since on-shell amplitudes, where all gluons have the same helicity, vanish.
In eq.~(\ref{on_shell_recursion}) the shifted spinors
$\hat{k}_A^1$, $\hat{k}_{\dot{B}}^n$, $\hat{K}_A$ and $\hat{K}_{\dot{B}}$ are given by
\bq 
 \hat{k}_A^1 = k_A^1 - z k_A^n, 
 & &
 \hat{K}_A = \frac{K_{A\dot{B}} k_1^{\dot{B}}}{\sqrt{\left\l 1+ \left| K \right| n+ \right\r}},
 \nonumber \\
 \hat{k}_{\dot{B}}^n = k_{\dot{B}}^n + z k_{\dot{B}}^1,
 & &
 \hat{K}_{\dot{B}} = \frac{k_n^A K_{A\dot{B}}}{\sqrt{\left\l 1+ \left| K \right| n+ \right\r}},
\eq
where
\bq
 K_{A\dot{B}} = \sum\limits_{l=1}^{j-1} k_A^l k_{\dot{B}}^l, 
 & & 
 K_{1,j-1}^2 = \mbox{det} \; K_{A\dot{B}},
\eq
and
\bq 
 z & = & \frac{K_{1,j-1}^2}{\left\l 1+ \left| K \right| n+ \right\r}.
\eq
The recurrence relation starts with $n=3$. The only non-vanishing amplitudes are 
\bq
A_3\left( k_A^1, k_{\dot{B}}^1, -, k_A^2, k_{\dot{B}}^2, -, k_A^3, k_{\dot{B}}^3, + \right)
 & = &
 i \sqrt{2} \frac{\l 1 2 \r^4}{\l 12 \r \l 23 \r \l 31 \r},
 \nonumber \\
A_3\left( k_A^1, k_{\dot{B}}^1, +, k_A^2, k_{\dot{B}}^2, +, k_A^3, k_{\dot{B}}^3, - \right)
 & = &
 i \sqrt{2} \frac{[ 2 1 ]^4}{[ 32 ] [ 21 ] [ 13 ]},
\eq
plus the ones with cyclic permutations of the helicities.
It should be noted that due to the particular choice of shifting the spinors, the three-point function with
$\hat{k}^1_A$ vanishes if the helicities are a cyclic permutation of $(-,-,+)$.
Similarly, the three-point function involving $\hat{k}^n_{\dot{B}}$ vanishes if the helicities are a cyclic
permutation of $(+,+,-)$.
To speed up the computation the Parke-Taylor formulae in eq.~(\ref{Parke_Taylor}) may be used for $n \ge 4$.

As in the previous method we have to update at each step in the recursion the list of Weyl spinors and the helicities.

\section{Performance and numerical stability}
\label{sect:numeric}

\subsection{Performance}

We have implemented all four methods into numerical programs.
For an unbiased comparison of the efficiencies of the different methods,
each author has programmed all four methods independently, in order
to eliminate possible dependencies on the programming skills of the programmer.
It turned out that all programs gave the same pattern in the 
study of efficiency and accuracy.

All methods give identical results within an accuracy of $10^{-12}$ 
for randomly chosen non-exceptional phase space points and up to $12$ external particles.
To investigate the performance in terms of CPU time 
we study the quantity ${\cal M}_n$ defined in eq.~(\ref{benchmark_M}):
\bq
 {\cal M}_n & = &
 \sum\limits_{\lambda_1,...,\lambda_n}
 \left| A_{n}\left( k_{1}^{\lambda_{1}}, ..., k_{n}^{\lambda_{n}} \right)
 \right|^2.
 \eq
It is clear from the algorithms that the first two methods (Berends-Giele and scalar diagrams)
need a constant amount of CPU time for each helicity configuration, whereas the last two methods
(MHV and BCF) are very efficient if the helicities are predominately all plus or all minus,
but take more CPU time if the helicity configuration contains roughly the same number of plus and minus
helicities.
To compare the different methods, the quantity ${\cal M}_n$ sums over all helicity configurations.
This corresponds to the situation encountered in the calculation of cross-sections and observables.
Table \ref{table:timing} shows the CPU time needed for the computation of ${\cal M}_n$
as $n$ varies from $4$ to $12$.
\begin{table}
\begin{center}
\begin{tabular}{|l|rrrrrrrrr|}
\hline
 $n$  & 4 & 5 & 6 & 7 & 8 & 9 & 10 & 11 & 12 \\
\hline
 Berends-Giele & 0.00005 & 0.00023 & 0.0009 & 0.003  & 0.011 & 0.030 & 0.09 & 0.27 & 0.7  \\
 Scalar        & 0.00008 & 0.00046 & 0.0018 & 0.006  & 0.019 & 0.057 & 0.16 & 0.4  & 1 \\
 MHV           & 0.00001 & 0.00040 & 0.0042 & 0.033  & 0.24  & 1.77  & 13   & 81  & --- \\
 BCF           & 0.00001 & 0.00007 & 0.0003 & 0.001  & 0.006 & 0.037 & 0.19 & 0.97  & 5.5 \\
\hline
\end{tabular}
\caption{\label{table:timing}
CPU time in seconds for the computation of the $n$ gluon amplitude on
a standard PC (2 GHz Pentium IV), summed over all helicities.
}
\end{center}
\end{table}
The test was done on a standard PC with a 2 GHz Pentium IV processor. 

As can be seen from the table, the Berends-Giele type recurrence relation is the fastest method, as
the number of external gluons increases.
In second place comes the method with scalar diagrams. 
As already discussed in the presentation of the algorithms, these two methods are fast due to the fact that
they can work with a static list of four-momenta and helicities.
This avoids copying large amounts of data at each step of the recursion.
The scalar diagram technique allows for a higher degree of optimisation in the subroutines,
but this is out-weighted by the fact that
in the Berends-Giele method each three- or four-valent vertex is called exactly once, whereas in the scalar diagram
method each vertex is called three times with different helicity configurations.
\begin{table}
\begin{center}
\begin{tabular}{|l|rrrrrrrr|}
\hline
 $n$  & 13 & 14 & 15 & 16 & 17 & 18 & 19 & 20 \\
\hline
 Berends-Giele & 2 & 4 & 11 & 27  & 64 & 149 & 367 & 831 \\
 Scalar        & 3 & 6 & 15 & 36  & 85 & 195 & 465 & 1043  \\
\hline
\end{tabular}
\caption{\label{table:timing_large_n}
Continuation of table \ref{table:timing} for $n$ in the range from $13$ to $20$ 
for the Berends-Giele method and the scalar diagram method.
The settings are as in table \ref{table:timing}.
}
\end{center}
\end{table}
Table \ref{table:timing_large_n} shows the timings 
for the Berends-Giele method and the scalar diagram method
for the computation of ${\cal M}_n$
as $n$ varies from $13$ to $20$.
It should be noted that for $n=20$ the results of the two methods agree within $10^{-11}$.
It can be seen from tables \ref{table:timing} and \ref{table:timing_large_n} that the time required by the 
scalar diagram method
grows slower than that required by the Berends-Giele method as the number $n$ of external particles increases.

The MHV method is rather slow compared to the other three methods. This is related to the double sum appearing in
eq.~(\ref{MHV_recursion}), which explicitly over-counts each contribution.
In addition, the look-up tables we used to speed up the calculation are in this case rather memory-intensive.
That is the reason why we were not able to compute the $12$-point amplitude within this approach.

The BCF method is faster than the Berends-Giele method as long as the number of external particles is below $9$.
For applications to three- or four-jet rates at LHC the BCF recurrence relations are therefore
an improvement in efficiency.
 
\subsection{Numerical stability}

\begin{figure}
\includegraphics[scale=0.7]{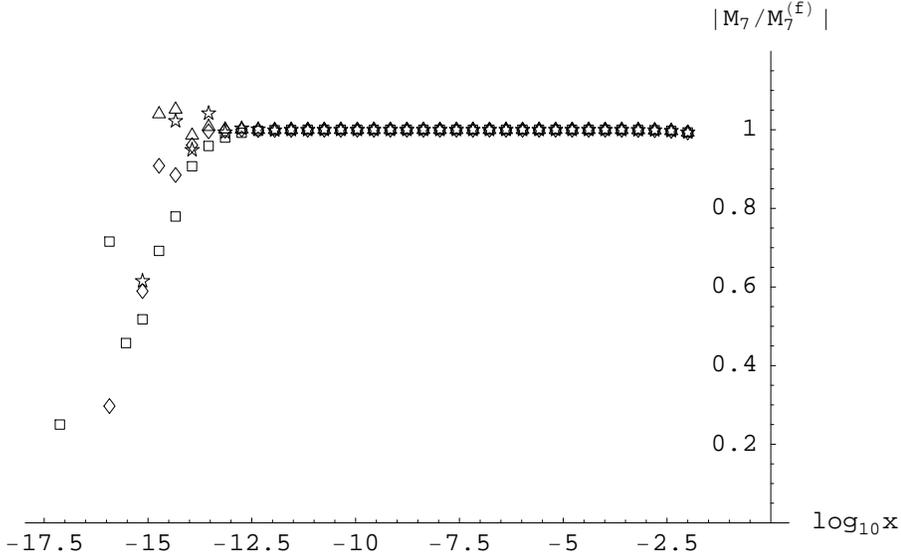}
\caption{\label{fg:soft} Ratio of the sum of squared helicity amplitudes to its factorised form for a set of 7-gluon configurations where
one gluon becomes soft.  $x$ is the energy fraction of the soft gluon.  
Key: $\diamond$ Berends-Giele $\star$ scalar diagrams $\triangle$ MHV rules $\Box$ on-shell.}
\end{figure}

\begin{figure}
\includegraphics[scale=0.7]{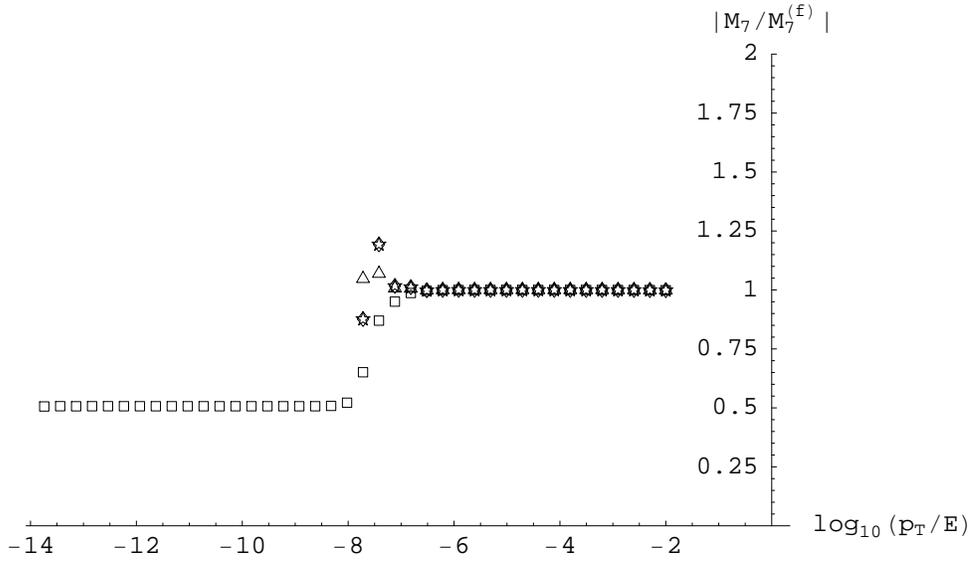}
\caption{\label{fg:coll} Ratio of the sum of squared helicity amplitudes to its factorised form for a set of 7-gluon configurations where
two gluons becomes collinear.  $p_T/E$ is the transverse momentum of the
pair of gluons, normalised to the total energy.  Key: $\diamond$ Berends-Giele $\star$ scalar diagrams $\triangle$ MHV rules $\Box$ on-shell.}
\end{figure}

We have already mentioned that all methods give identical results for
randomly chosen non-exceptional phase space points 
within an accuracy of $10^{-12}$.
In this section we study the numerical stability near exceptional phase space points,
e.g. near singular configurations where one or more partons become unresolved.
We limit ourselves to single unresolved configurations, where 
an external momentum becomes soft,
or two external momenta become collinear.
In these cases the quantity ${\cal M}_n$ exhibits an infrared singularity and factorises into a singular
function and a lower-point amplitude, 
as described by the equations (\ref{eq:soft_singularity}) and (\ref{eq:coll_singularity}).
The singular behaviour can cause problems with the numerical
stability of amplitude calculations.  To investigate this problem, we evaluated
${\cal M}_n$ for configurations approaching
each kind of singular limit.  To illustrate the stability, we have plotted in
Figures \ref{fg:soft} and \ref{fg:coll} the
ratio of ${\cal M}_7$ to its factorised form ${\cal M}_7^{(f)}$ as given
by the right-hand sides of eqs.~(\ref{eq:soft_singularity}) and (\ref{eq:coll_singularity}).
The soft limit (Fig.~\ref{fg:soft}) is described by $x \to 0$ where $x$ is the
fraction of the total energy carried by the soft gluon.  The onset of instability is at $x \simeq 10^{-12}$.
The collinear limit 
(Fig.~\ref{fg:coll}) is described by $p_T/E \to 0$ where $p_T$ is
the transverse momentum involved in the collinear splitting (with $z=1/2$).
Instability occurs when the dimensionless variable $p_T/E \simeq 10^{-7}$.
We observe no significant differences between the four methods.

\begin{figure}
\includegraphics[scale=0.7]{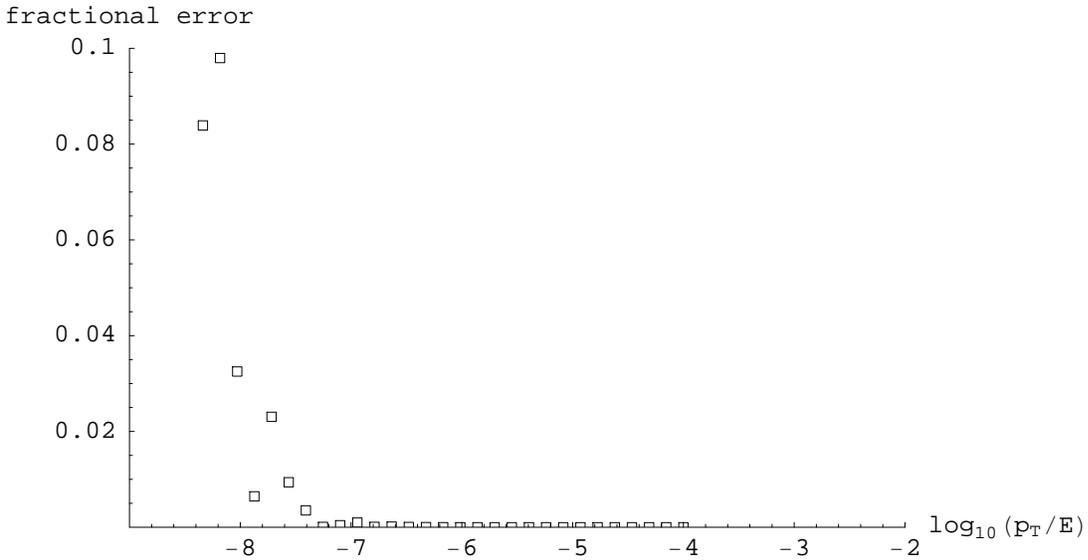}
\caption{\label{fg:spur} Fractional error in the sum of squared helicity amplitudes 
computed with the on-shell recursion relations for a set of 6-gluon configurations where
$k_1 + k_2$ becomes collinear to $k_3 + k_4$.  $p_T/E$ is the transverse momentum between
the 2 pairs of gluons, normalised to the total energy.  }
\end{figure}

In addition to these physical singularities, spurious singularities might occur.
An example can be found in the on-shell recursion relation.
The shift in the spinors introduces sandwiches of the form
$\l p_i- | p_k+p_l | p_j- \r$ in the denominator.
For example, the analytical formula for the six-gluon partial amplitude
$A_6(1^+,2^+,3^+,4^-,5^-,6^-)$ reads:
\bq
\label{eq:spurious}
\lefteqn{
A_6(1^+,2^+,3^+,4^-,5^-,6^-) = } & & 
 \nonumber \nopagebreak \\
 & &
 4 i \left[
           \frac{\l 6- | 1+2 | 3- \r^3}{\l 61 \r \l 12 \r [ 34 ] [ 45 ] s_{126} \l 2- | 1+6 | 5-\r}
         + \frac{\l 4- | 5+6 | 1- \r^3}{\l 23 \r \l 34 \r [ 56 ] [ 61 ] s_{156} \l 2- | 1+6 | 5- \r}
     \right].
\eq
This introduces
unphysical singularities when sums of external momenta
become collinear.  Of course, these cancel exactly in the final result, but can lead
to problems when the recursion relation is evaluated numerically.  
An example
of this is shown in Figure \ref{fg:spur}.  Here we consider an amplitude of the form
shown in eq.~(\ref{eq:spurious}), in the limit that $p_1+p_6$ becomes collinear to
$p_2+p_5$.  We have
plotted the fractional error in the on-shell results for the sum over helicity amplitudes 
by comparing to those of the Berends-Giele
recursion relation.  The onset of instability occurs when
the transverse momentum is of the order of
$10^{-7}E$.

\begin{figure}
\includegraphics[scale=0.7]{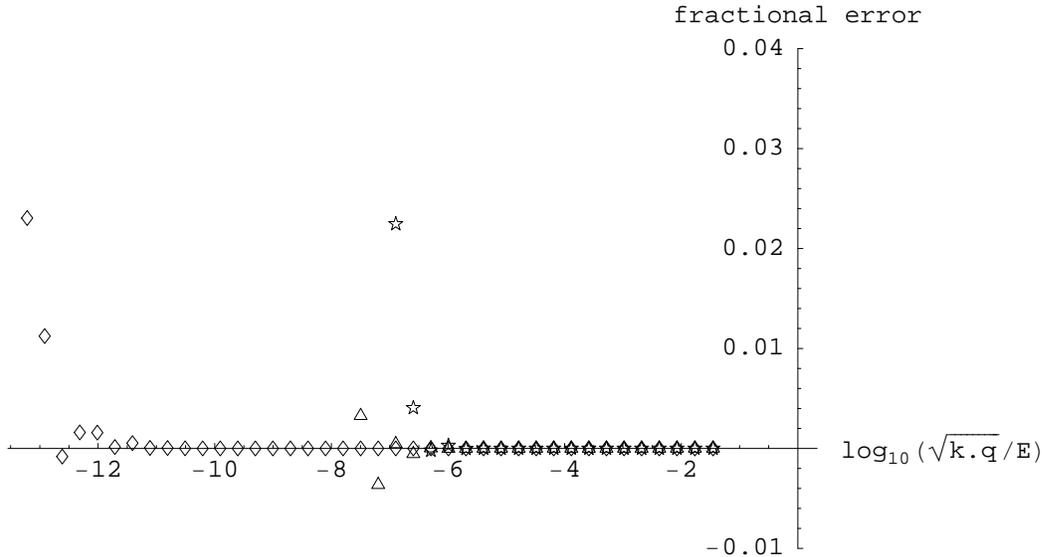}
\caption{\label{fg:ref} Fractional error in the sum of squared helicity amplitudes 
as the reference vector ($q$) used in the definition of each recursion relation becomes
collinear with an external momentum ($k$). Key: $\diamond$ Berends-Giele $\star$ scalar diagrams $\triangle$ MHV rules.
  }
\end{figure}

The other recurrence relations can also exhibit spurious singularities, as each require an
arbitrary light-like ``reference'' vector $q$ to be specified, and various quantities diverge
if this vector becomes collinear to one of the external momenta.  For the Berends-Giele
recurrence relations this vector is needed to define the polarisation vectors in eq.~(\ref{eq:pol}),
and for the scalar diagram and MHV vertex approaches it is needed to fix the on-shell
projection in eq.~(\ref{offshellcont}).  The dependence of our results on $q$ as $q$ becomes
collinear to an external momentum $k$ is illustrated in Fig.~\ref{fg:ref}.
The scalar and MHV results become unstable
when $\sqrt{k.q} \simeq 10^{-7}E$, whereas the Berends-Giele recurrence relation is stable down to
$\sqrt{k.q} \simeq 10^{-12}E$.
This behaviour is expected, since in the Berends-Giele recurrence relation the reference vector $q$
enters only the external polarisation vectors, whereas
in the other two methods it also affects the internal lines.


Overall, all methods exhibit satisfactory numerical stability properties.
As far as spurious singularities are concerned, the Berends-Giele method performs slightly better.

\section{Conclusions}
\label{sect:conclusions}

In this paper we studied numerical implementations of recursive methods for the computation of 
Born gluon amplitudes.
These amplitudes (together with corresponding ones, where additional quarks or vector bosons are involved),
are relevant for LHC physics. They enter numerical LO or NLO program codes.
As these calculations are based on Monte-Carlo integration over the phase space, the efficiency of the computation
has a direct impact on the running time of the Monte-Carlo program.

From the four methods considered, we found the Berends-Giele method performs best, as the number of external
partons increases ($n \ge 9$).
However, for a not so large number of external partons ($n<9$), the on-shell recursion relation
offers the best performance.
As this is the range most relevant to LHC physics, this new method leads to an improvement
in the numerical computation of Born amplitudes.

We also investigated the numerical stability and accuracy. Here, all methods give satisfactory
results.


\begin{appendix}

\section{Spinors}
\label{appendix:spinors}

For the metric we use
\bq
g_{\mu \nu} & = & \mbox{diag}(+1,-1,-1,-1).
\eq
We define the light-cone coordinates as
\bq
p_+ = p_0 + p_3, \;\;\; p_- = p_0 - p_3, \;\;\; p_{\bot} = p_1 + i p_2,
                                         \;\;\; p_{\bot^\ast} = p_1 - i p_2.
\eq
In terms of the light-cone components of a null-vector, the corresponding massless spinors $\l p \pm |$ and $| p \pm \r$ 
can be chosen as
\bq
\left| p+ \right\r = \frac{e^{-i \frac{\phi}{2}}}{\sqrt{\left| p_+ \right|}} \left( \begin{array}{c}
  -p_{\bot^\ast} \\ p_+ \end{array} \right),
 & &
\left| p- \right\r = \frac{e^{-i \frac{\phi}{2}}}{\sqrt{\left| p_+ \right|}} \left( \begin{array}{c}
  p_+ \\ p_\bot \end{array} \right),
 \nonumber \\
\left\l p+ \right| = \frac{e^{-i \frac{\phi}{2}}}{\sqrt{\left| p_+ \right|}} 
 \left( -p_\bot, p_+ \right),
 & &
\left\l p- \right| = \frac{e^{-i \frac{\phi}{2}}}{\sqrt{\left| p_+ \right|}} 
 \left( p_+, p_{\bot^\ast} \right),
\eq
where the phase $\phi$ is given by
\bq
p_+ & = & \left| p_+ \right| e^{i\phi}.
\eq
Spinor products are denoted as
\bq
 \l p q \r = \l p - | q + \r = p^A q_A,
 & &
 [ q p ] = \l q + | p - \r = q_{\dot{A}} p^{\dot{A}}.
\eq 
We will also use the notation
\bq
 \l p \pm | k | q \pm \r & = & \l p \pm | k_\mu \gamma^\mu | q \pm \r.
\eq


\section{Feynman rules}
\label{appendix:Feynman}

\subsection{Colour-ordered Feynman rules}
\label{appendix:colour_ordered_rules}

The Feynman rules for colour-ordered partial amplitudes read:
\bq
\begin{picture}(100,35)(0,55)
\Vertex(50,50){2}
\Gluon(50,50)(50,80){3}{4}
\Gluon(50,50)(76,35){3}{4}
\Gluon(50,50)(24,35){3}{4}
\LongArrow(56,70)(56,80)
\LongArrow(67,47)(76,42)
\LongArrow(33,47)(24,42)
\Text(60,80)[lt]{$k_{1}$,$\mu$}
\Text(78,35)[lc]{$k_{2}$,$\nu$}
\Text(22,35)[rc]{$k_{3}$,$\rho$}
\end{picture}
 & = &
 i \left[
          g^{\mu\nu} \left( k_1^\rho - k_2^\rho \right)
        + g^{\nu\rho} \left( k_2^\mu - k_3^\mu \right)
        + g^{\rho\mu} \left( k_3^\nu - k_1^\nu \right)
   \right],
 \nonumber \\
\begin{picture}(100,75)(0,50)
\Vertex(50,50){2}
\Gluon(50,50)(71,71){3}{4}
\Gluon(50,50)(71,29){3}{4}
\Gluon(50,50)(29,29){3}{4}
\Gluon(50,50)(29,71){3}{4}
\Text(72,72)[lb]{$\mu$}
\Text(72,28)[lt]{$\nu$}
\Text(28,28)[rt]{$\rho$}
\Text(28,72)[rb]{$\sigma$}
\end{picture}
 & = &
   i \left[
        2 g^{\mu\rho} g^{\nu\sigma} - g^{\mu\nu} g^{\rho\sigma} 
                                     - g^{\mu\sigma} g^{\nu\rho}
 \right].
 \nonumber \\
\eq


\subsection{Scalar diagrammatic rules}
\label{appendix:scalar_rules}

The non-vanishing primitive vertices involving only gluons are:
\bq
\begin{picture}(100,35)(0,55)
\Vertex(50,50){2}
\Gluon(50,50)(50,80){3}{4}
\Gluon(50,50)(76,35){3}{4}
\Gluon(50,50)(24,35){3}{4}
\Text(60,80)[lt]{$1^-$}
\Text(78,35)[lc]{$2^-$}
\Text(22,35)[rc]{$3^+$}
\end{picture}
 & = &
 V_3(k_1^-,k_2^-,k_3^+) 
 =
 i \sqrt{2} \l 1 2 \r \frac{[3q]^2}{[1q][2q]}
 =
 i \sqrt{2} \frac{\l 1 2 \r^4}{\l 12 \r \l 23 \r \l 31 \r},
 \nonumber \\
\begin{picture}(100,75)(0,50)
\Vertex(50,50){2}
\Gluon(50,50)(50,80){3}{4}
\Gluon(50,50)(76,35){3}{4}
\Gluon(50,50)(24,35){3}{4}
\Text(60,80)[lt]{$1^+$}
\Text(78,35)[lc]{$2^+$}
\Text(22,35)[rc]{$3^-$}
\end{picture}
 & = &
 V_3(k_1^+,k_2^+,k_3^-) 
 =
 i \sqrt{2} [ 2 1 ] \frac{\l 3 q \r^2}{\l 1 q \r \l 2 q \r}
 =
 i \sqrt{2} \frac{[ 2 1 ]^4}{[ 32 ] [ 21 ] [ 13 ]},
 \nonumber \\
\begin{picture}(100,75)(0,50)
\Vertex(50,50){2}
\Gluon(50,50)(71,71){3}{4}
\Gluon(50,50)(71,29){3}{4}
\Gluon(50,50)(29,29){3}{4}
\Gluon(50,50)(29,71){3}{4}
\Text(72,72)[lb]{$1^+$}
\Text(72,28)[lt]{$2^+$}
\Text(28,28)[rt]{$3^-$}
\Text(28,72)[rb]{$4^-$}
\end{picture}
 & = &
 V_4(k_1^+,k_2^+,k_3^-,k_4^-) 
 \nonumber \\
 & = &
 i \frac{[1q][2q]\l 3 q \r \l 4 q \r}{\l 1 q \r \l 2 q \r [3q] [4q]}
   \left( 
          1 + \frac{\left\l q- \left| 2-3 \right| q- \right\r \left\l q- \left| 4-1 \right| q- \right\r}
                   {\left\l q- \left| 2+3 \right| q- \right\r \left\l q- \left| 4+1 \right| q- \right\r} 
   \right),
 \nonumber \\
\begin{picture}(100,75)(0,50)
\Vertex(50,50){2}
\Gluon(50,50)(71,71){3}{4}
\Gluon(50,50)(71,29){3}{4}
\Gluon(50,50)(29,29){3}{4}
\Gluon(50,50)(29,71){3}{4}
\Text(72,72)[lb]{$1^+$}
\Text(72,28)[lt]{$2^-$}
\Text(28,28)[rt]{$3^+$}
\Text(28,72)[rb]{$4^-$}
\end{picture}
 & = &
 V_4(k_1^+,k_2^-,k_3^+,k_4^-) 
 \nonumber \\
 & = &
 i \frac{[1q] \l 2q \r [3q] \l 4 q \r}{\l 1 q \r [2q] \l 3 q \r [4q]}
   \left( 
          \frac{\left\l q- \left| 1-2 \right| q- \right\r \left\l q- \left| 3-4 \right| q- \right\r}
               {\left\l q- \left| 1+2 \right| q- \right\r \left\l q- \left| 3+4 \right| q- \right\r} 
 \right. \nonumber \\
 && \left.
        + \frac{\left\l q- \left| 2-3 \right| q- \right\r \left\l q- \left| 4-1 \right| q- \right\r}
               {\left\l q- \left| 2+3 \right| q- \right\r \left\l q- \left| 4+1 \right| q- \right\r} 
          - 2
   \right).
 \nonumber \\
\eq

\section{Splitting functions}
\label{appendix:splitting functions}

In the collinear limit the all-gluon tree-level partial amplitudes factorise according to
\bq
A_{n+1}(...,k_a,k_b,...) \stackrel{k_a || k_b}{\longrightarrow} \sum\limits_{\lambda = \pm} \mbox{Split}_{-\lambda} 
(k_a^{\lambda_a},k_b^{\lambda_b}) A_n(...,K^\lambda,...),
\eq
where $k_a$ and $k_b$ are the momenta of two adjacent legs.
In the collinear limit we have
$K = k_a + k_b$, $k_a = z K$ and $k_b = (1-z) K$. $\lambda$, $\lambda_a$ and $\lambda_b$ denote the
corresponding helicities.
The splitting functions are:
\begin{eqnarray}
\mbox{Split}_{g^+}(g^+,g^+) = 0, & & \mbox{Split}_{g^-}(g^-,g^-) = 0, \nonumber \\
\mbox{Split}_{g^+}(g^+,g^-) = \sqrt{2} \frac{(1-z)^{\frac{3}{2}}}{\sqrt{z} \langle a b \rangle }, & & 
 \mbox{Split}_{g^-}(g^-,g^+) = - \sqrt{2} \frac{(1-z)^{\frac{3}{2}}}{\sqrt{z} [ a b]}, \nonumber \\
\mbox{Split}_{g^+}(g^-,g^+) = \sqrt{2} \frac{z^{\frac{3}{2}}}{\sqrt{(1-z)} \langle a b \rangle }, & & 
 \mbox{Split}_{g^-}(g^+,g^-) = - \sqrt{2} \frac{z^{\frac{3}{2}}}{\sqrt{(1-z)} [ a b]}, \nonumber \\
\mbox{Split}_{g^+}(g^-,g^-) = - \sqrt{2} \frac{1}{\sqrt{z(1-z)} [ a b ] }, & & 
 \mbox{Split}_{g^-}(g^+,g^+) = \sqrt{2} \frac{1}{\sqrt{z(1-z)} \langle a b \rangle}.
\eq

\section{Optimising techniques}
\label{appendix:optimization}

In this appendix we comment briefly on optimisation techniques we employed to speed up the computation.
These techniques apply to all the four methods discussed in the main text.
Recursive formulations have the disadvantage that they evaluate the same quantity over and over again.
To overcome this obstacle, look-up tables are employed.
C++ already offers in the standard library the data structure
\verb/std::map/, which is well suited for look-up tables.
The template takes the types of three classes as arguments
\begin{verbatim}
 std::map< class_key, class_result, class_compare > look_up_table;
\end{verbatim}
where \verb/class_key/ represents a class, which stores the input information on which a result depends.
\verb/class_result/ defines the data type of the results and \verb/class_compare/ is a class, which provides
an operator 
\begin{verbatim}
 class class_compare
   {
   public:
     bool operator() (const class_key & a, const class_key & b) const;
   };
\end{verbatim}
for comparing two instances of type \verb/class_key/, which is needed to keep the look-up table
sorted.
The class \verb/std::map/ needs of the order $\log(n)$ operations to look-up a specific entry, if the table
is filled with $n$ entries.
This can be improved by using a method with constant look-up time.

A second point concerns temporary variables. All methods operate on data structures, like four-vectors or spinors, 
which are rather time-consuming if they need to be copied.
C++ allows operator overloading and for example the addition of fourvectors can be coded as follows
\begin{verbatim}
fourvector p,q;
fourvector sum = p + q;
\end{verbatim}
However, this first calculates the sum of $p$ and $q$ in a temporary variable and copies the result in a second step
to \verb/sum/.
It is more efficient for selected operations to write a method
\begin{verbatim}
 void add(fourvector & sum, const fourvector  & p, const fourvector & q);
\end{verbatim}
which avoids copying temporaries:
\begin{verbatim}
fourvector sum,p,q;
add(sum,p,q);
\end{verbatim}
This is not as elegant as in the first code fragment, but more efficient when used in low-level routines.
\end{appendix}


\end{document}